# IONS IN FLUCTUATING CHANNELS: TRANSISTORS ALIVE

Bob Eisenberg

Department of Molecular Biophysics and Physiology

Rush University and Medical Center

1653 West Congress Parkway

Chicago IL 60612 USA

Mathematics and Computer Science Division

Argonne National Laboratory

9700 S. Cass Avenue

Argonne, IL 60439 USA

# Abstract

Ion channels are proteins with a hole down the middle embedded in cell membranes. Membranes form insulating structures and the channels through them allow and control the movement of charged particles, spherical ions, mostly $Na^+$, $K^+$, $Ca^{++}$, and $Cl^-$. Membranes contain hundreds or thousands of types of channels, fluctuating between open conducting, and closed insulating states. Channels control an enormous range of biological function by opening and closing in response to specific stimuli using mechanisms that are not yet understood in physical language. Open channels conduct current of charged particles following laws of Brownian movement of charged spheres rather like the laws of electrodiffusion of quasi-particles in semiconductors. Open channels select between similar ions using a combination of electrostatic and 'crowded charge' (Lennard-Jones) forces. The specific location of atoms and the exact atomic structure of the channel protein seems much less important than certain properties of the structure, namely the volume accessible to ions and the effective density of fixed and polarization charge. There is no sign of other chemical effects like delocalization of electron orbitals between ions and the channel protein. Channels play a role in biology as important as transistors in computers, and they use rather similar physics to perform part of that role. Understanding their fluctuations awaits physical insight into the source of the variance and mathematical analysis of the coupling of the fluctuations to the other components and forces of the system.

Transistors are everywhere in our life, so widespread that the younger generation hardly knows they exist. The singular importance of transistors is hidden nowadays in the millions of FETs that remember our snapshots. The importance of transistors was obvious to everyone when radios contained just four [1].

Transistors are the vital elements of our electronic technology because they amplify and switch so well according to the simple laws of electrodiffusion. In a semiconductor switch, only a few hundred holes or electrons are needed to switch or control signals of billions of charges every billionth of a second in devices so small that they can easily be held on our wrist or even in our ear. Transistors switch quickly because the mass of the quasi-particles that carry electric charge in them is so small. Holes and 'electrons' have very little mass or inertia and take little energy to accelerate[*].

Transistors are not the only tiny elements that control current flow in our wrist or ear. While physicists and engineers were creating transistors in germanium and silicon, biophysicists—who might be called channologists—were discovering life's transistors in biological cells. These analogs of transistors are specialized proteins that control electricity (and much else) in the biological tissues and cells of our wrist or ear. Life's transistors are ion channels.

## 1. What are Ion Channels?

Ion channels are proteins with a hole down their middle (Fig. 1) that provide a controllable path for electrodiffusion of ions through biological membranes (Fig. 2). It seems reasonable to assume that ions in channel proteins move much as they do in bulk solution, although that is surely a working hypothesis to be attacked and modified as we learn more and more. Electrodiffusion in solutions follows nearly the same simple laws as electrodiffusion in semiconductors [2-4] even though current in solutions is carried by real, not quasi-particles. Current flow in water solutions, and ion channels, is carried by spherical ions chiefly $Na^+$, $K^+$, $Ca^{++}$, and $Cl^-$. (The charge on these ions is permanent in the sense that it does not change under biological conditions.)

Ion channels are found in biological membranes that surround and define biological cells. Membranes without channels are nearly perfect insulators preventing DC current flow; membranes provide the insulating and isolating substrate that allows channels to control the flux of ions, current, and electricity into cells, much as $SiO_2$ provides an insulating and isolating substrate for transistors.

Nearly all biology occurs in cells (Fig. 2) [5]. Ion channels control flows in and out of cells and so an enormous range of cellular life is controlled by these proteins in health [6] and disease [7]: in that sense, ion channels are the transistors of life, controlling life nearly as completely as transistors control solid state devices and machines [8-13].

Ion channels are used to control most functions of cells because they act as gatekeepers for cells, providing paths for the movement of ions and messages in and out of cells, in particular, controlling (nearly) all the electrical properties of cells and tissues. Information processing and signaling in the nervous system use electrical signals controlled by channels; sensory organs make electrical signals using channels; contraction of voluntary (skeletal) muscle and cardiac muscle is controlled by electrical signals and channels. The heart functions as a pump because its contraction is coordinated by channels. Kidneys, lungs, stomach, intestine, endocrine glands, sweat glands use channels to transport substances—you name your tissue of interest, except red blood cells, evidently. Diseases strike channels and the study of 'channelopathies' is one of the fastest growing areas of medicine [8,9]. (Search for 'channelopathy' on the internet to see what I mean.) Thousands of molecular biologists study channels every day, manipulating the channel protein (or its DNA blueprint) with the magnificent tools of molecular biology, recording current through single channel molecules using the reconstitution and patch clamp methods of Nobel laureates Sakmann and Neher [14]. Hundreds of structural biologists map the location of individual atoms of channel proteins, thanks to Nobel Laureate Rod MacKinnon, more than anyone else. Channologists form a significant fraction of all biologists because channels play such a significant role in health and disease.

Why do ion channels have such an important role in biology? Channels have come to be important as part of the evolutionary process that created them; and evolution is a chaotic process (in the mathemati-cal sense of the word), reset by random catastrophes at stochastic intervals. It is not clear that enough can ever be known in hindsight to reconstruct the trajectory of evolution of a specific channel because a complete description of a process at a single time is often not enough to regenerate (i.e., 'determine' in the language of mathematics) the previous history (i.e., trajectory in time) of that process, particularly if the process, like evolution, has many of the properties of a backwards heat equation [15,16], involves many interacting variables and has a complex stochastic history. For that reason, even exhaustive experimentation may well be insufficient to understand how channels came to do what they do today.

---

[*]I wish the negative quasi-particles of silicon/germanium were named (something like) semiconductor-electrons—with 'semi-electrons' as a nickname—so they are not confused with isolated electrons that flow in a vacuum. Surprisingly few scientists are aware that the negatively charged quasi-particles of silicon/germanium are not the isolated electrons of physics textbooks and cathode ray tubes.

But I believe we can know enough to understand how channels work, and to manipulate and control them, even if we cannot understand how they came to work that way. I believe we can understand channels much as we understand complex inanimate devices, much as we now understand transistor devices.

Transistors work because engineers and physicists built a structure providing a useful current voltage relation, that follows a simple input output relation, when power is supplied to drive holes and semi-electrons through them. Ion channels work because evolution built a structure and used particular physics to drive ions through them, providing a useful current voltage relation, that follows 'laws' (input output relations) just being discovered in the last few years.

Channels work (mostly) by their fluctuating opening and closing. The holes down the middle of channels switch stochastically from closed to open to closed forming a random telegraph signal. The open probability (i.e., duty cycle) of the channel controls the total charge movement (i.e., integrated current of ions) across the membrane. Each of the thousands of types of channels has a different controller of its duty cycle; each has different types of gates that respond to different types of signals. Some channels respond to chemical signals of a molecule or two, others respond to mechanical stretch, still others, respond to electrical potential itself. Engineers are trying to make channel devices that exploit their special sensitivity, hoping channels are no harder to handle in a technological environment than the soap films of our liquid crystals, LCD monitors and TV sets. Interestingly, the mathematics of liquid crystals [17-20] may prove to be the most useful mathematics for ions in channels [21-23].

## 2. Gating in Channels and Transistors: different physics.

The gating process of channels has an analogous role to the gating process of transistors, but it does not have analogous physics. Channels use gating motions that involve mass and friction. Transistors do not. The gating of transistors does not involve substantial movement of mass but rather depends on changes in the shape of the electric field. An analogy between gating in transistors and channels [10,12] confuses the essentially different physics of opening and closing in the two devices. The physics is different not because one system is physical and one is biological, but because changing the electric field and changing the location of mass are different, in whatever context the change occurs. The physics is different also because of the state of our knowledge. We have essentially complete understanding of gating in transistors over the entire range of scales from macroscopic function to atomic structure. We do not have adequate knowledge of the gating mechanism of channels. Many biologists are working on gating, but agreement on even the structural basis of gating is not yet at hand. It is not wise to make a physical model of a system of unknown structure. At least in biology it is better to await the rapid progress of our structural colleagues.

## 3. Current flow in Open Channels and Conducting Transistors: a useful analogy.

The analogy between current flow in a transistor and an open channel is good physics despite these differences—once the channel is open, after the channel protein has finished its conformation changes [11,13]. Indeed, current flow in transistors and the open channel follow nearly the same mathematical laws because the current flow of ions and quasi-particles is governed by nearly the same physics.

Ions and quasi-particles move under the control of gradients of concentration and electric potential. The paths of holes and semi-electrons can be ballistic. The paths of ions are never ballistic. They are (more or less) the trajectories of mathematical Brownian motion. The theory of Brownian motion has been used to analyze their motion and to construct a rigorous theory of diffusion as a chemical reaction [24-33].

Electric potential plays a particularly important role when these laws are applied to channels because the channels are so small. The pores of ion channels are from 4 to (say) 9 Å in diameter, and the control regions of channels are thought to be only a few Angstroms long. The pores are so small that only a few elementary charges carried by a few ions are enough to produce substantial potentials; pores have tiny capacitance. Potentials in these pores are important also because the potential scale of biology is small; cell membranes are lipid films, formed of two layers of lipid molecules only some 30 Å thick—think of soap bubbles or films of olive oil thin enough to form black films on still water—and so breakdown is seen at potentials of hundreds of millivolts. (The reader should work out the field strength to see why.) Most of life's processes and most of channel function occurs at potentials smaller than 200 mV; indeed control occurs at potentials of 1–2 mV, much smaller than the thermal potential of 25 mV $= k_b T/e$ under biological conditions. Thus, the location and nature of electric charge have a large role in controlling channels and biological function.

Biological pores typically contain a handful of permanent charges in their walls. These charges reside in the atoms of the amino acids that make up proteins, and play a role quite analogous to the role of doping in transistors. In the ordinary case, these permanent charges do not change value as ions move through open channels. The charges also do not change position, if position is measured in averages on the biological time scale of μsec and longer, although the positions certainly fluctuate a great deal on the atomic scale of the speed of sound (see p. 845 of [3]). How these charges move and change, as the channel gates, as proteins change conformation, or as proteins do chemistry, making and breaking covalent bonds, is an important area of future physical investigation. Indeed, I have long suspected that generation and recombination of 'permanent' charges of amino acids—in protein biochemistry called protonation and deprotonation of acidic and basic residues—play a crucial role in the function of transport molecules closely related to channels [13]

The physics of ion motion in channels is the physics of electrodiffusion much as it is in transistors. Electricity and diffusion interact much as they do in liquid crystals [22,34]. Diffusion moves charge, charge changes the electric field. The equations of electricity and diffusion must be solved together, just as they are in computational electronics. The diffusion field of ions is created by the difference in the concentration of ions inside and outside cells. These concentrations are described by inhomogeneous Dirichlet boundary conditions (different concentrations at different places) that inject mass, and energy into the channel. The electric field is created by different types of charge: the charge of other ions, the permanent charges of the protein, the induced (polarization) charge on molecules and at interfaces, and the charge on electrodes and in surrounding baths. The charged surface of proteins is an inhomogeneous Neumann boundary condition: the jump in normal derivative of the potential (weighted by the different dielectric coefficients) is set by the permanent charge on the boundary.

The surfaces of the proteins are not maintained at fixed potentials. They are not connected to sources of charge. On the other hand, the electrodes on either side of the membrane are typically maintained at different fixed potentials and so form Dirichlet boundary conditions that inject mass, energy, and current into the system.[†] Channel systems are necessarily far from equilibrium when they function as devices because their function is (usually) to conduct current.

Equilibrium thermodynamic analysis of a device (or channel) is usually not helpful, if the goal is to understand and control it. The function of devices has little to do with their thermodynamics and so thermodynamics tells little about how devices work or can be controlled: devices do not work at thermodynamic equilibrium, i.e., when their power inputs are all connected to the same zero potential.

### 4. Channels like transistors are fluctuating devices.

The analysis of devices is in many ways the core of engineering and is really quite different from the analysis of general physical systems. Devices have a purpose, usually summarized in an input output law, valid only under a limited set of conditions. The goal of studying devices is to understand and manipulate that input output law and so it is rarely worth studying devices under general conditions. It is particularly useless to study conditions in which devices do not work (unless one is interested in studying the failure mode of devices).

Devices in biology can be defined by similar sentences, although it is important to define 'purpose' more precisely and objectively. As any physiologist or physician can tell you, the purpose of a an organ, tissue, cell, or cell component is its input output relation. The purpose of the heart is to pump blood according to an input output relation; the purpose of cardiac muscle is to shorten so the heart can pump; the purpose of channels in cardiac muscle is to initiate and coordinate the contraction of the cardiac muscle, and so on. The purpose of each structure in cardiac muscle is clear: the purpose is to provide a definite output for a given input that can be used by other structures to sustain the life of the animal (so it can survive and reproduce, if one wishes to reach all the way to evolutionary biology in our discussion). In favorable cases, these input output relations in biology can be written quantitatively and objectively as equations or computer programs. The purpose of biological devices is no more vague and subjective than the purpose of an amplifier.

---

[†]In biological cells, active processes using chemical energy maintain average potential and concentration across membranes. Signaling in nerve and muscle fibers involves transient changes in electrical potential but the potential during the time between signals is maintained constant in healthy cells. In experiments, specialized apparatus, made of transistors, maintains and controls these variables.

The input output relations in biological systems often form a hierarchy of scales, with smaller devices providing outputs needed by larger devices for the overall function of the cell, tissue, or organ. In the case of nerve fibers, and cardiac muscle to a lesser extent, one can write and solve equations across almost the entire length scale from molecules to macroscopic function. These are the device equations of the biological system and I believe the purpose of the biological system is to execute these design equations in nearly the same sense that the purpose of a typical amplifier is to multiply a voltage by a constant.

Device equations tell how to use an amplifier; thermodynamics does not. Device equations describe current-voltage relations of transistors. Device equations must have spatially inhomogeneous boundary conditions if the input and output of the device are to be distinguished. The goal of much of biology, as of engineering is to design and control devices, not to study every possible property of the device. Thus, the boundary conditions that control the device and keep it working properly must be included in the analysis. Analysis of differential equations with spatially uniform boundary conditions, or with boundary conditions defined vaguely at infinity, cannot easily describe the inputs and outputs of devices and so is of limited use when dealing with biological or engineering systems.

Scientists have only begun to discover the 'device equations' that describe the input output relations of channels. We seek equations that tell us how the potential and concentrations far from channels control their function. The electric potential outside the channels, in baths and on electrodes, can be measured but the potential inside channels is not known. The electric potential in proteins can only be calculated from the equations defining the electric field. These equations depend on all charge and so must include all the charges present. The electric field is produced by charges, but it also exerts force on charges and changes their location in an important way, called shielding or screening. Thus, the value of the electric field changes significantly with experimental conditions.

Shielding plays a very important role in determining the electrical properties of systems with mobile charge, in many cases dominating those properties [35-37]. Proteins are like that. All the charge in proteins—permanent and mobile and induced (i.e., polarization charge created by the electric field)—creates potential, but potential fields move only some charge. The moved charge screens permanent charges and has a dramatic effect on the net effect of those permanent charges.

## 5. Chemical Kinetics and Open Ion Channels.

Proteins have usually been described in the tradition of chemical kinetics. The binding and transport properties of proteins—as well as the chemical reactions in which proteins participate—are traditionally described by rate constants independent of concentration, ionic strength, and other conditions [5,6,38,39]. Rate constants of ionic systems cannot be independent of concentration [33,40]. Rate models describe current through open channels as the movement of ions over a potential barrier of constant size independent of conditions [6,38]. Thus, the barriers (and rate constants) in traditional rate models of proteins are immune from the effects of screening/shielding that determine many of the physical properties of systems of mobile charge [35-37].

Misrepresenting potential profiles as constants, independent of conditions, is particularly serious, [33,40] because it implies the injection of charge and energy into the system just at its most sensitive place, at the peak of potential barriers, where function is controlled. When conditions change, the only way to maintain a profile of potential is to inject charge in many places along that profile. That injection is artificial and serious changes the system. Models that represent channels or proteins as potential profiles (or rate constants) independent of conditions are artificial and distort the system by injecting charge as conditions where no charge is actually injected in the real world.

Models with this defect are unlikely to provide much insight into function. This failure of the chemical tradition to deal with the fundamental properties of the electric field is a significant source of the difficulties scientists have in calculating drug binding and protein function and folding, in my opinion, although not necessarily in the opinion of others. If my view is correct, no amount of computer resources will resolve these problems until the electric field is dealt with in a calibrated way, i.e., in a way shown to give the macroscopic results measured in simple systems [41-43]

In a physical analysis, current flow through open channels must be computed by a combination of Poisson and transport equations so that the electric field that moves charge—and is in itself changed when charge moves—can be computed self-consistently. The equations must be solved together to predict fields, much as current flow is analyzed in transistors in computational electronics, because transport changes charge, charge changes potential, and potential changes transport [41-43]. The potential landscape of a protein or channel indeed determines the forces on ions and substrates—and determines protein function, drug binding, and so on—but the potential landscape must be calculated from all the charges present, including at the boundaries, and must be recalculated every time charge moves, as is done in computational electronics, with atomic resolution in space (Å) and time (femtoseconds).

## 6. Brownian Motion of Charged Particles, limitations in Einstein's analysis.

The implications of this statement for statistical physics are profound, as they are for biophysics, both at equilibrium and in general. Langevin equations of Brownian motion always require Poisson's equation in this view if particles have significant charge anywhere on their surface.[‡] Transport changes charge densities, charge changes potential, and potential changes transport, whether we work at the macroscopic or atomic scale of resolution, and so the equations of transport and electric field must be solved together, and they must be solved including boundary conditions, and are hard to treat with periodic boundary conditions, if they are different at different locations. Einstein's and Langevin's equation (in which ink particles move randomly in an electric field, even if the field is zero and therefore not shown explicitly) must always be coupled to a Poisson equation so the fluctuating field can be computed from the charges and their fluctuating position.

If the Brownian motion is calculated in a mean electric field, as Einstein and Langevin did, the calculation does not describe the actual random motion of charged particles, which occurs in fluctuating fields. Estimates of variance are obviously wrong in such calculations. Estimates of means may also be wrong because Langevin equations coupled to Poisson are very nonlinear processes. The mean value of such processes can depend on the fluctuations of the underlying noise. Indeed, qualitative properties of coupled Langevin-Poisson processes are likely to be quite different from the qualitative properties of the mean field Langevin systems studied by Einstein and Langevin. Ions of one type—in mixed solutions containing other types of ions—may move against their own gradient of electrochemical potential if the electric field driving their migration is dominated by other ions, for example, those present at much larger number densities.

The Einstein/Langevin treatment of diffusion also does not allow flow, if it is used in the high friction Smoluchowski limit, with a Maxwellian distribution of velocities [25]. We use the phrase 'Maxwellian distribution' in a strict sense here. In this case, the mean velocity of the Maxwellian is zero always for all conditions. The Maxwellian in this definition is symmetrical and thus its velocity and flow are identically zero. This Maxwellian is incompatible with boundary conditions that force flow. If the macroscopic flow is zero, the average velocity must be zero. Local equilibrium cannot produce global nonequilibrium by mathematics alone. Something must be done to perturb the local equilibrium of the Maxwellian and that must be present in a revised Maxwellian that describes a perturbed local equilibrium that allows flow. We have shown [25], by mathematics alone, that the change in the Maxwellian is tiny but essential.

The change arises from a revision of the Einstein/Langevin/Smoluchowski treatment. Einstein/Langevin/Smoluchowski uses a first order differential equation and so it cannot accommodate different boundary conditions at different places. Diffusion systems commonly involve two baths with two different controlled concentrations. A first order differential equation cannot describe these two concentrations as boundary conditions. The Einstein/Langevin/Smoluchowski treatment cannot accommodate two differ-ent boundary conditions and so it cannot accommodate macroscopic flow driven by two different concentrations that are kept at two different values by boundary conditions.

---

[‡] The ink particles that Brown and Perrin studied and Einstein and Langevin described are charged colloids. Note that a water molecule is highly charged locally even though its global charge is zero and the field created by this charge extends many diameters, well beyond the repeat distance used in most simulations of water or proteins that employ periodic boundary conditions.

For the same reason the Einstein/Langevin/Smoluchowski treatment cannot be used for a system with inputs and outputs. Systems with inputs and outputs clearly require two boundary conditions to specify the different properties at different locations (that define input and output). Ion channels and transistors have inputs and outputs, as do almost all devices. Ion channels and transistors have flows driven by gradients of concentration imposed by different boundary conditions. In other words the Einstein/Langevin/Smoluchowski treatment cannot deal with the most commonplace systems of electrochemistry or cell physiology. Those systems all have distinct inputs and outputs and different properties at different places.

### 7. Brownian Treatment of Macroscopic Flux Requires a Second Order Stochastic Differential Equation.

The derivation of device equations from stochastic differential equations in a system with distinct inputs and outputs requires careful mathematics, particularly in the high friction limit [25]. Spatially nonuniform boundary conditions (that produce nonzero average velocity and flow) can be combined with the high friction limit only if a more general treatment is used in which the first order stochastic differential equation of the Einstein/Langevin/Smoluchowski treatment is replaced with a second order stochastic differential equation. The velocity distribution which is the solution of this second order stochastic differential equation does not have zero mean. Indeed, its mean is precisely the mean velocity of particles corresponding to the macroscopic flux.

The friction that appears in this second order stochastic differential equation is very large. The large value can be exploited in an analysis in which the velocity is preserved as a variable: asymptotic analysis and singular perturbation theory are used to exploit the large value of the friction while approximating the second order stochastic differential equation we call the full Langevin equation. The result is very pleasing [25]. The distribution of velocities needed to produce macroscopic flux is not a Maxwellian, but rather a Maxwellian displaced by a constant value. Every velocity is displaced by the same amount and that amount is exactly what is needed to account for the flux. The displaced Maxwellian is the result of the high friction. The displaced Maxwellian is the statistical signature of the drift diffusion equation of computational electronics. Local equilibrium is incompatible with macroscopic flux, but a displaced Maxwellian can describe both local and global diffusion, both Brownian motion and macroscopic diffusion from bath to bath, with exact mathematical consistency. The displaced Maxwellian should replace the Maxwellian of local equilibrium in the stochastic treatments of diffusion, in my opinion. Otherwise, inconsistencies are inevitable. Different research groups are likely to deal with the inconsistencies in different ways producing confusing results. Everyone gets the same results in the analysis of transistors and semiconductors in computational electronics where the displaced Maxwellian is always used.

### 8. Electrodiffusion in Computational Electronics.

Transistors and semiconductors are analyzed by computational electronics, one of the most successful of the computational sciences. Computational electronics calculates the properties of transistors with essentially no adjustable parameters, a striking accomplishment in multi-scale analysis. Computational electronics starts with the atomic properties of matter and successfully calculates the macroscopic currents by which transistors function on long time scales. This computational success over an enormous range of scales is one of the main reasons electronic and semiconductor technology has been so successful. This multiscale success is what is sought in computations of ionic solutions and proteins but—I think it fair to say—is not yet in hand.

The treatment of the electric field and electrodiffusion in computational electronics is strikingly different from their treatment in ionic solutions and proteins and one must suspect that the difference has something to do with the relative success of the fields in computing useful macroscopic properties. The focus in computational electronics is on the electric field and the flow of current. It is taken for granted that the field and flow must be computed 'to infinity'. The field computation must include the boundaries where power is supplied by different voltages at different places. The calculations must include spatially inhomogeneous boundary conditions. Periodic boundary conditions do not easily accommodate such conditions, particularly 'at infinity' and so are (essentially) never used in computational electronics [41-43]. In compu-tational electronics, care is taken to describe the electric field over all space and time, even if some atomic detail must be sacrificed. In computational chemistry and biology care is taken to describe atomic detail, even if the long range properties of concentration and electric field must be sacrificed.

Computational electronics computes the electric field in this way because understanding devices requires such computation. It was apparent from the beginning that any model of a transistor must include the value of the voltage applied to its leads. It is obvious to an engineer that devices cannot execute their device equations without power supplies and so devices can only be understood if their analysis includes different boundary potentials at different locations. After all, anyone who has built a device containing FETs knows the importance of the potential applied to transistor terminals. A FET can be many different devices depending on the voltages applied to it, and the engineer chooses the device he wishes by adjusting the values of the power supply voltages. It is obvious that these voltages must be included in theory, if the different devices are to be defined, let alone understood. What is not obvious, but is in fact true, is that even low resolution equations of computational electronics describe transistors quite well, with a single set of parameters, if those equations include spatially inhomogeneous boundary conditions, power supplies, and flow. The key is to understand the electric field including, of course, the sources that produce it.

The nonequilibrium properties of the device do not have to be described in much detail in most device equations because flux usually arises from the spatial non-uniformity of boundary conditions—not from complex properties of the differential operators. The differential operators are the same whether the device is turned off, at equilibrium, with spatially uniform boundary conditions, or in operation, with spatially nonuniform boundary conditions. The differential operators describing the physical model of devices are the same whether the power supply is present or not. The essential properties of devices are seen even in low resolution models, in which the velocities fall into a simple Maxwellian distribution displaced by a constant, which is the mean velocity, the flux in different units, in fact.

Computational electronics had the insight from its very beginning that current flow in semiconductor solids—whatever its physical mechanism—should be described in the tradition of device analysis. Computational electronics described current flow in semiconductor solids as the consequence of the mean electric field applied to terminals, just as current flow was described in vacuum tubes [44, particularly p. 65, 11, and 144) ]. I suspect this approach seemed so natural to the founders of semi-electronics that it was nearly unconscious, but whatever the historical reason, this insight is remarkable and is not used in computational chemistry or biology. In computational chemistry and biology, current flow and electric fields are sometimes not computed at all, and certainly do not have a central place.

**9. The electric field dominates.**

The novelty and significance of the treatment of semiconductor devices should not be forgotten, just because it is now usual, taught to millions of students each year. Everything in our semiconductor technology depends on this insight that the electric field dominates and must be computed and understood in general, from transistor terminal to terminal, including the spatially nonuniform potentials and current flow that make transistors work.

This approach to the analysis of electrodiffusion grew naturally from the analysis of vacuum tubes (or valves as they were more aptly called in the mother tongue of English engineers). Viewed naively, electrons in a vacuum and (quasi-particle) electrons in a solid semiconductor do not seem similar. It is certainly not clear that they should follow similar transport laws. Nonetheless, electrons and semi-electrons are similar, and follow similar laws, and so transistors could be built using the experience of vacuum tube design, starting first with the description of the mean electric field created by the steady potentials applied by power supplies.

Computational electronics says it is the electric field that matters, more than anything else. Devices with similar electric fields behave in (qualitatively) similar ways, no matter how the fields were created, no matter what carries the current (within reason). Mimic the electric fields of a triode, and you will have an amplifier and switch, no matter where the fields are created, if anything flows in those fields in a reasonable way.

Mimic the fields of a vacuum or semiconductor diode in a protein and you will have a rectifying channel. That is an unmistakable prediction to a computational engineer. It remains the task of the channologists to check that prediction and find its limitations. We have been trying [45]. But the analyses of electrodiffusion must include flow and spatially nonuniform boundary potentials, as well as Langevin/Poisson equations, if they are to describe devices, as well as electrodiffusion. Only in the last decade or so have channologists realized [46-49] that the principles and tools of computational electronics can be used to understand the rectification of current flow through open channels studied in detail since the time of Hodgkin and Huxley, the 1950s, and glimpsed much earlier, nearly one hundred years ago.

This then is the proper and useful analogy between transistors and channels. Transistors alive are the open channels of cell membranes; once open, channels and transistors both follow the same laws of electrodiffusion [11,13]

Of course, the analogy between semi-electron and hole flow and ionic current is not complete. The electrical property of rectification is not the only or the most important property of ions in solution or in open channels. Proteins and ions have chemical properties that quasi-particles lack; and computational chemistry must join computational electronics if the resulting chemical properties of channels are to be understood.

### 10. Chemical properties of ionic channels: selectivity.

The chemical properties of ion channels and proteins are of great interest both for historical and scientific reasons. Historically, the great majority of workers in molecular biology have been trained in chemistry, not in physics or electronics; only a few of us were lucky enough to be trained both by molecular biologists and channologists. Thus, the chemical properties of proteins are described on nearly every page of any textbook of biochemistry or molecular biology, but even the most elementary discussion of electricity is not found there. (Search for a dielectric constant, or any equation at all, in textbooks of biochemistry, if you wish to check this sweeping statement.)

Biologists study the chemical properties of channels and proteins because they are so striking. Channels and proteins, for example, select between different chemicals (e.g., drugs) with great specificity; channels respond selectively to ions that differ only a little in diameter or charge, with otherwise identical chemical properties [5-7]. The chemical selectivity arising from channels has long been considered one of the special characteristics of life.

It was natural to believe, as I did for decades, that the special chemical selectivity of channels arose from special chemistry. I thought that selectivity came from 'chemical interactions' between ions and special binding sites on proteins, designed by evolution to bind ions specifically [39]. It seemed natural to describe chemical interactions in the tradition of chemical kinetics, as chemical reactions, involving delocalization of electrons in the outer orbits of the atoms of the protein, requiring the solution of Schrödinger's equation one way or another [38]. But these ideas did not work very well. No one was able to design and build selective channels using this chemical tradition.

### 11. Selectivity from Physics, diameter, and charge

Chemical specificity can arise another way. Chemical specificity can arise from physical factors, not involving delocalization of outer electrons, not involving binding sites with specific atomic geometry. In highly concentrated solutions, for example, the free energy per mole of $Na^+$ and $K^+$ are quite different, even though the ions differ only in diameter. Modern physical chemistry shows that the energy necessary to crowd spheres together in large concentrations depends a great deal on the diameter and charge of the spheres [2,4]. The main chemical property of such solutions (the free energy per mole usually called 'the activity') is determined by the diameter and charge of these spheres, much more than by anything else. The special chemical properties of water, the hydration shells around ions, and other chemical phenomena, are not involved, to a first order, or even second order, except as they determine the dielectric properties of the concentrated salt solution. If the number density, diameter, and dielectric properties of the salt solution are known, the free energy per mole can be calculated accurately without regard to other chemical properties of the solution.

In this view, concentrated salt solutions are viewed as compressible plasmas; the (volume of the) solution itself is incompressible, but (the number densities of) its components are not. The number densities of components of the solution vary a great deal and that variation determines many of the properties of the solution, even though the gravimetric density of the solution is nearly invariant.

Computational chemistry has given us a computational theory of selectivity in concentrated salt solutions. The question is whether this theory is relevant to ions in channels. The answer is that the theory is relevant if ions in channel proteins are concentrated salt solutions.

Ions in a channel protein are highly concentrated because proteins in general—and active sites and channel pores in particular—'bristle with charge' [50]. The large density of permanent charge on the walls of ion channels, and on the active sites of proteins, guarantees a large concentration of ions nearby: deviations from electric neutrality must be small, even in the tiny structures of proteins and channels because proteins can only tolerate small voltages.

The number densities of ions in channels are enormous. The L-type calcium channel, which controls the contraction of the heart, and is the target of the 'calcium channel blockers' used by many of our physicians, has four permanent negative charges in its active site. The four mobile positive charges nearby have a number density of some 30 molar, $\sim 2 \times 10^{22}$ cm$^{-3}$. The charges are very crowded indeed (water is ~ 55 molar; solid NaCl ~37 molar). I conclude that physical effects, calculated with physical theories and simulations, are enough to understand the biological property of selectivity of L type calcium channels [51-55].

The question then is to find the role of the protein among these physical effects: what is the role of the channel protein in this combination of computational chemistry and electronics? How does the protein, and ultimately the genome, control selectivity?

In our view [51-55], the answer is that the structure (and charge distribution) of the protein guarantees the existence of crowded charge; physics controls the energy of those charges.

The channel protein determines selectivity in much the same way that an engine block determines the properties of an automobile motor. In one sense, the engine block does little. Its job is to hold things in place and not to move. In another sense, the engine block does everything. If the engine block warps even a tiny amount $(\sim 10^{-5})$, pistons seize up, and the motor dies.

In this view of selectivity, developed over the years in work with Wolfgang Nonner, the channel protein provides the structure for selectivity, just as the engine block provides the structure for the automobile motor. The channel protein provides the permanent charge and dielectric charge, in the right place; it provides mechanical strength. The channel protein controls selectivity much as an engine block controls combustion. Both provide the arena in which physics and chemistry provide the energy that drives the machine. In this view, the protein should be viewed as a solid machine built at considerable cost, which stores free energy, and is not in a configuration of minimum free energy, any more than an engine block or amplifier is itself in a configuration of minimum free energy. A channel protein in this view is a device, a simple kind of machine, not a complex chemical system at or near equilibrium.

In this view, the channel protein does not delocalize electrons to provide a binding site. Rather, it produces a binding site by determining the permanent charge and volume of its pore, much as an automobile engine controls piston function by determining the diameter and strength of the cylinders in which pistons slide back and forth.

This physical view of selectivity is very different from that of structural biology, where biological function (and selectivity) is treated as the direct consequence of chemical bonds between atoms seen in x-ray crystallography of crystals of proteins.

In the physical view, selectivity is mostly a physical consequence of the small size and large charge density of active sites of channels and proteins. Of course, selectivity in the binding of asymmetrical molecules (like drugs) will involve the static shape of the drug molecules and their permanent electric fields, and selectivity will probably depend on the induced charge (i.e., field dependent polarization) of their electrons. It is also clear that this physical view of selectivity is only a working hypothesis to be believed only as it is tested: one must keep looking for evidence of the role of chemical bonds in selectivity.

Working this way, I think one can approach biology much as one approaches engineering or physics. One can make a specific model and refine and improve it by adding more effects, more physics, and quantum chemistry, step by step, as needed. The approach of twentieth century biology, using verbal models, or reaction schemes from gas phase chemical kinetics, seems less likely to succeed, however poetic the words or complex the schemes, in my opinion. The approach of the twenty-first century using simulations in atomic detail is more likely to succeed, in my view. Atomic detail simulations will become more useful [56] as they are refined [57] to compute the activity of concentrated salt solutions over the range of biological importance from μM to many molar.

**12. Transistors Alive**

We have come a long path in considering transistors alive. Transistors have an important analog in life, ion channels. Ion channels control much of biological function, as transistors control technology. The physics of control of transistors and channels are quite different, while other properties of ion channels and transistors are quite similar. Electrodiffusion controls the motion of ions in channels, once they are open, much as it controls the motion of the quasi-particles, holes and semi-electrons, in semiconductors. But holes and electrons are not ions. Ions have size and chemical properties that holes and semi-electrons lack and so computational chemistry must be combined with computational electronics to understand the chemical selectivity of ions crowded into channels by the electric field. Simple models of crowded charge do surprisingly well as models of selectivity in one highly selective biological channel. The combination of computational chemistry and computational physics should lead to a biotechnology of channels as important to industry—and more important to medicine and our daily life—as the electro-technology of semiconductors.

**Captions**

Fig. 1. A chemist's view of ionic channels. The vertices of the line segments represent atoms, whose locations have been determined by diffraction analysis of x-ray scattering from crystals of the protein. The surfaces are more or less surfaces of constant electrical potential, in a qualitative computation. Two different channels are shown, at right angles to each other. The hole down the middle is filled with a mixture of water molecules and ions (not shown), which conduct electrical current. The ions are at very high number density.

Fig. 2. A textbook author's view of channels in a biological cell. The membrane of the cell is an insulating structure in which channel proteins are embedded that allow and control the movement of charged particles, spherical ions, mostly $Na^+$, $K^+$, $Ca^{++}$, and $Cl^-$. Open channels conduct current of charged particles following laws of electrodiffusion rather like the laws of electrodiffusion of quasiparticles in semiconductors. Channels control an enormous range of biological channel by opening and closing: many types of channels are present in membranes, most of which are closed at any moment. For both these reasons, channels can be said to be life's transistors.

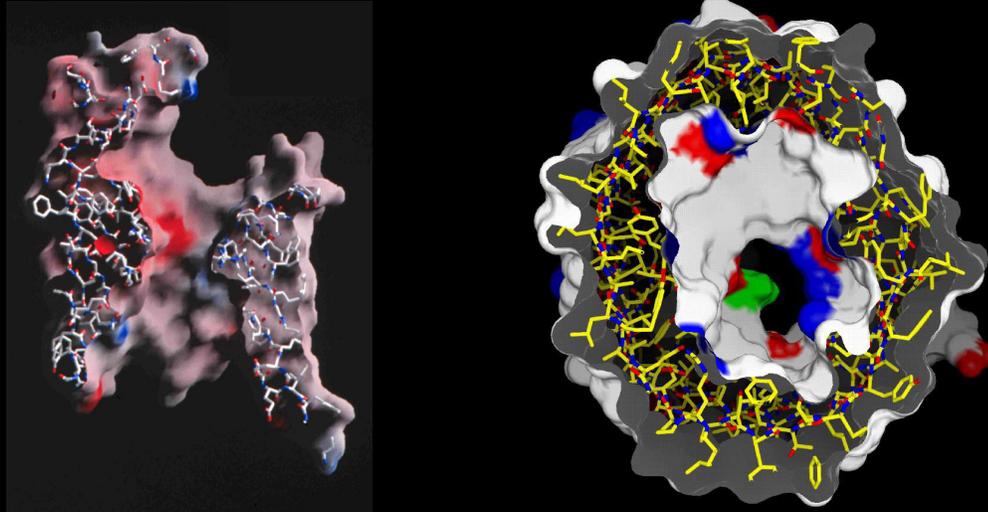

Fig. 1. A chemist's view of ionic channels. The vertices of the line segments represent atoms, whose locations have been determined by diffraction analysis of x-ray scattering from crystals of the protein. The surfaces are more or less surfaces of constant electrical potential, in a qualitative computation. Two different channels are shown, at right angles to each other. The hole down the middle is filled with a mixture of water molecules and ions (not shown), which conduct electrical current. The ions are at very high number density.

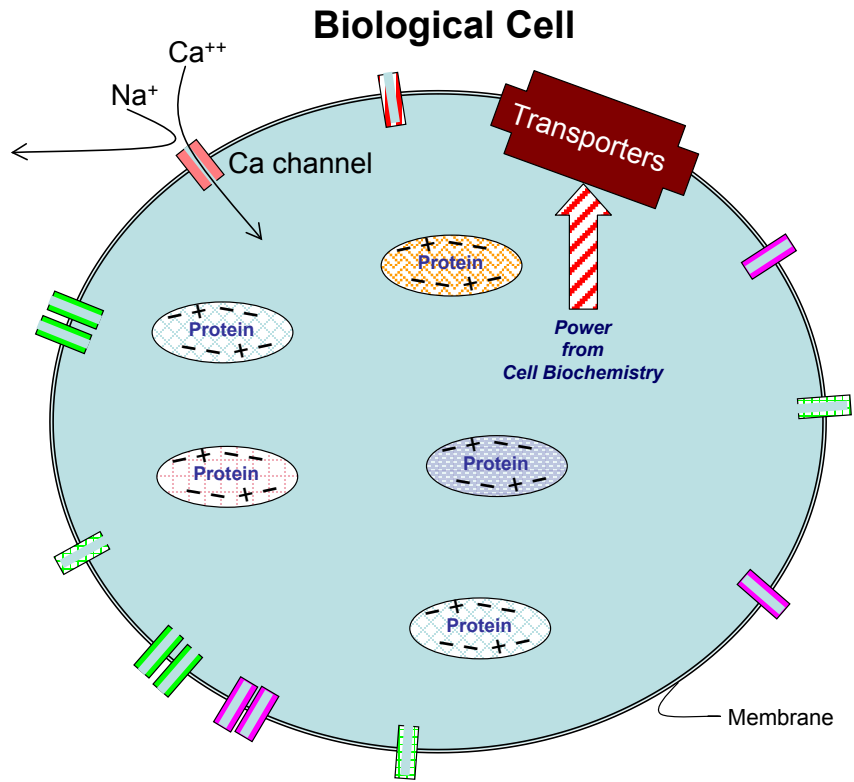

Fig. 2. A textbook author's view of channels in a biological cell. The membrane of the cell is an insulating structure in which channel proteins are embedded that allow and control the movement of charged particles, spherical ions, mostly $Na^+$, $K^+$, $Ca^{++}$, and $Cl^-$. Open channels conduct current of charged particles following laws of electrodiffusion rather like the laws of electrodiffusion of quasiparticles in semiconductors. Channels control an enormous range of biological channel by opening and closing: many types of channels are present in membranes, most of which are closed at any moment. For both these reasons, channels can be said to be life's transistors.